\documentclass[amsmath,amssymb,floatfix,pra,reprint]{revtex4-1}
\usepackage{graphicx}
\usepackage{bm}
\usepackage{amsmath,amssymb}
\usepackage{color}
\usepackage{gensymb}
\usepackage[caption=false]{subfig}
\usepackage{enumitem}
\usepackage{amsmath}
\usepackage{fullpage}
\usepackage{hyperref}
\usepackage{float}
\usepackage{times}
\usepackage{soul}

\hypersetup{
    colorlinks = true,
    allcolors = [rgb]{0,0,0},
}

\newcommand{\gamp}{\Gamma}
\newcommand{\gamr}{\gamma}
\newcommand{\gampower}{\gamma_c}
\newcommand{\kp}{k}
\newcommand{\kc}{k_\mathrm{c}}
\newcommand{\s}{5S_{1/2}}
\newcommand{\pstate}{5P_{3/2}}
\newcommand{\dstate}{5D_{5/2}}
\newcommand{\sexcited}{7S_{1/2}}
\newcommand{\vT}{v_\mathrm{T}}

\begin{document}
\raggedbottom
\title{Power narrowing: Counteracting Doppler broadening in two-color transitions}
\author{Ran Finkelstein}
\thanks{These authors contributed equally to this work}
\affiliation{Department of Physics of Complex Systems, Weizmann Institute of Science, Rehovot 760001, Israel}

\author{Ohr Lahad}
\thanks{These authors contributed equally to this work}
\affiliation{Department of Physics of Complex Systems, Weizmann Institute of Science, Rehovot 760001, Israel}

\author{Ohad Michel}
\affiliation{Department of Physics of Complex Systems, Weizmann Institute of Science, Rehovot 760001, Israel}

\author{Omri Davidson}
\affiliation{Department of Physics of Complex Systems, Weizmann Institute of Science, Rehovot 760001, Israel}

\author{Eilon Poem}
\affiliation{Department of Physics of Complex Systems, Weizmann Institute of Science, Rehovot 760001, Israel}

\author{Ofer Firstenberg}
\affiliation{Department of Physics of Complex Systems, Weizmann Institute of Science, Rehovot 760001, Israel}

\begin{abstract}
Doppler broadening in thermal ensembles degrades the absorption cross-section and the coherence time of collective excitations. In two photon transitions, it is common to assume that this problem becomes worse with larger wavelength mismatch. Here we identify an opposite mechanism, where such wavelength mismatch leads to cancellation of Doppler broadening via the counteracting effects of velocity-dependent light-shifts and Doppler shifts. We show that this effect is general, common to both absorption and transparency resonances, and favorably scales with wavelength mismatch. We experimentally confirm the enhancement of transitions for different low-lying orbitals in rubidium atoms and use calculations to extrapolate to high-lying Rydberg orbitals. These calculations predict a dramatic enhancement of up to 20-fold increase in absorption, even in the presence of large homogeneous broadening. More general configurations, where an auxiliary dressing field is used to counteract Doppler broadening, are also discussed and experimentally demonstrated. The mechanism we study can be applied as well for rephasing of spin waves and increasing the coherence time of quantum memories. \end{abstract}
\maketitle

Doppler broadening is ubiquitous in atomic and molecular spectroscopy. Atoms and molecules with different thermal velocities experience different Doppler shifts, which broaden the absorption lines and reduce their contrast \cite{Cohen-Tannoudji}. Increasing the light intensity typically causes further broadening due to saturation or other power-broadening mechanisms, such as inhomogeneous light-shifts. These broadening or dephasing mechanisms are major limiting factors, particularly in the field of quantum optics with atomic ensembles \cite{Hammerer2010,Huber2011,Ripka2018,Bromley2016,Kash1999}. For instance, the coherence time of collective excitations in atomic gasses is often limited by Doppler dephasing, hindering the performance of single-photon sources and memories \cite{finkelstein2018fast,Kaczmarek2018,Lee2016}. 
As Doppler broadening is an inhomogeneous dephasing mechanism, one can potentially counteract it by introducing additional velocity-dependent shifts \cite{Grynberg1979,CCT1979,popov2000enhanced,yavuz2013suppression,Lahad2019}. Here we study the counteraction of Doppler broadening by velocity dependent light-shifts and identify an important class of systems where such counteraction occurs naturally, without a need for additional auxiliary fields.

Coherent two-photon processes, such as Raman transitions, two-photon absorption, and electromagnetically induced transparency (EIT), are at the heart of many quantum-optics protocols, ranging from quantum light sources and memories \cite{Duan2001,Eisaman2005,Reim2011,Katz2018} to sensing and quantum non-linear optics \cite{firstenberg2016review}.   
The canonical example of a three-level system employed for these processes is the $\Lambda$ configuration, where two long-lived ground states are coupled via an intermediate excited state. In these systems, two-photon transitions are usually characterized by negligible residual Doppler broadening, as the nearly-degenerate transition wavelengths experience opposite Doppler shifts \cite{Firstenberg2013}. Particularly in a degenerate $\Lambda$ system under EIT conditions, atoms at all velocities contribute to slowing down light and storing it \cite{fleischhauer2005EIT}. Moreover, the coherence formed between the two lower states is usually long-lived, which further contributes to narrow linewidths. This class of systems can thus operate coherently at room temperature, making it a viable quantum technology platform \cite{Kitching2005}. 

\begin{figure}[t!]
\includegraphics[width=\columnwidth]{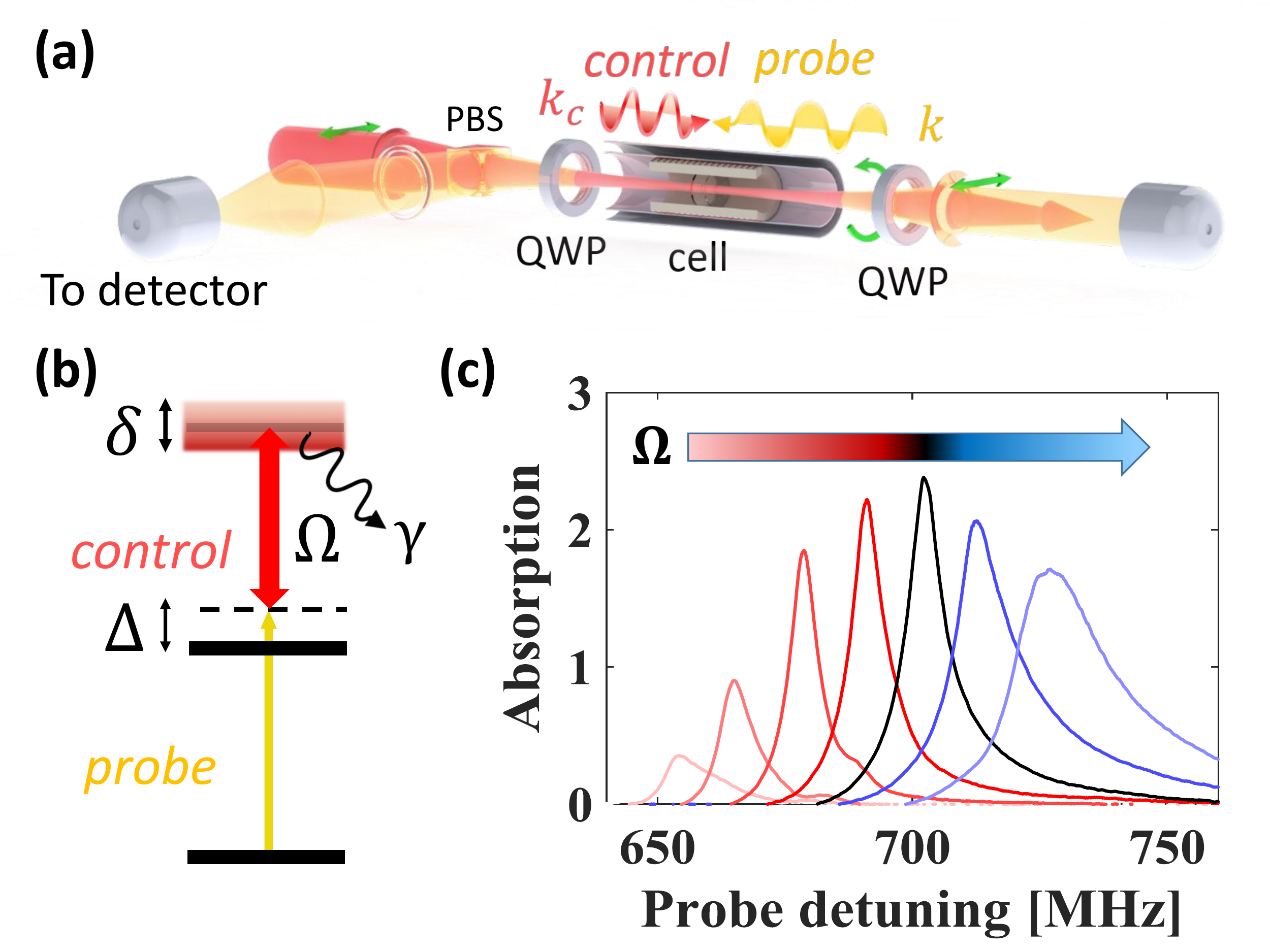}
\caption{\textbf{Spectroscopy of two-color transitions.} (a) \textit{probe} and \textit{control} beams with different wave-numbers $\kp \neq \kc$ counter-propagate inside a vapor cell. Both beams are circularly-polarized (using quarter-wave plates, QWP; polarizations shown by green arrows), and a polarizing beam-splitter (PBS) picks out the transmitted probe. (b) The ladder configuration, where wavelength mismatch introduces two-photon Doppler shifts $\delta(v)$ that depend on the atomic velocity $v$. The control field couples the intermediate state to a higher excited state with Rabi frequency $\Omega$ and detuning $\Delta$. (c) Measured two-photon absorption spectra ($\Delta=650$~MHz) exhibiting power narrowing. As the power is increased, the linewidth decreases and the absorption increases, up to an optimal power (black spectrum), after which the line broadens.} \label{FigSetup}
\end{figure}
 Another practical, though less studied, three-level system is the ladder configuration \cite{Grove1995}, where the intermediate state is coupled to an even higher excited state [Fig.~\ref{FigSetup}~(b)]. Ladder-type systems were recently shown to be advantageous for broadband, noise-free, photon storage \cite{finkelstein2018fast,Kaczmarek2018}, and they form the basis for quantum nonlinear optics with Rydberg atoms \cite{firstenberg2016review}. 
Nevertheless, a major caveat in ladder-type systems is the residual Doppler broadening due to wavelength mismatch of the two optical transitions, which leads to significant dephasing and broadening of the two-photon resonance \cite{Urvoy2013}. 

Here, we demonstrate and study power narrowing of two-color two-photon excitations in ladder systems, \emph{i.e.} the cancellation of residual Doppler broadening by light-shifts. We find the conditions under which the spectra of these systems present resonances with no motional broadening and high contrast. 
Power narrowing and the resulting line enhancement are observed for both two-photon absorption and EIT resonances. We provide a model which reproduces the observations and explains them in terms of velocity-dependent light-shifts, generated by the excitation light.

In order to place power narrowing in a broader context, we explore several extensions of the above study.
First, we identify the universal condition for optimal power narrowing and derive its scaling with the system parameters.
Second, we provide a prediction for two-color excitations of high-lying Rydberg states, which are typically characterized by a large wavelength mismatch. We show that power narrowing becomes even more significant in this case, and the resulting two-photon absorption is greatly enhanced.
Third, we establish power narrowing as a general effect of far-detuned dressing fields by demonstrating and analyzing power narrowing by means of an auxiliary dressing field.

\emph{System and model.---} We study the absorption spectrum of a weak probe beam, counter-propagating a strong control beam with Rabi frequency $\Omega$ in a thermal vapor cell, as depicted in Fig.~\ref{FigSetup}(a,b). The wave-numbers of the probe and control are $\kp$ and $\kc$, respectively. We define the unitless parameter \mbox{$q=\kc/\kp-1 $}, which quantifies the wavelength mismatch and thus the magnitude of the residual Doppler broadening $q\kp\vT$, where $\vT$ is the atomic thermal velocity. 
The homogeneous decoherence rates of the intermediate and excited states are $\gamp$ and $\gamr$, respectively. A two-photon resonance appears when the two-photon detuning $\delta$ is close to zero. It has the form of a transmission line due to EIT when the one-photon detuning $\Delta$ from the intermediate level is close to zero and an absorption line due to  two-photon absorption when $\Delta$ is large. The control power can be varied, and, for each power, the probe frequency is scanned to obtain an absorption spectrum [see Fig.~1(c)]. In a thermal ensemble, the one-photon resonances are broadened due to the Maxwell-Boltzmann distribution $f(v)=(2\pi \vT^2)^{-1/2} e^{-v^2/(2\vT^2)}$ of atomic velocities $v$. We consider the typical regime for hot atoms, where the resulting Doppler broadening dominates the linewidth  \mbox{$\kc\vT,\kp\vT \gg \gamp$}. For describing the three-level system, we first examine the far-detuned limit $\Delta\gg\gamp$, where resonant two-photon absorption occurs. We consider the vast majority of the atoms with velocities $v$ satisfying \mbox{$|\Delta-kv|\gg \gamp$} for which the absorption spectrum is given by \cite{fleischhauer2005EIT}
\begin{equation}
\label{Absoprtion_theory}
    \alpha(\delta,v)=\alpha_0\gamp\frac{\frac{\Omega^2}{(\Delta-kv)^2}\gampower}{\gampower^2+(\delta+qkv-\frac{\Omega^2}{\Delta-kv})^2},
\end{equation}
where $\gampower=\gamr+\gamp\frac{\Omega^2}{(\Delta+kv)^2} $ is the power-broadened width and $\alpha_0$ is the resonant absorption coefficient for stationary atoms. The two terms in the denominator that shift the two-photon resonance are the residual Doppler shift, associated with the two-photon momentum transfer $qk$, and the velocity-dependent light-shift induced by the control field. When integrating over all velocities $\alpha(\delta)=\int{dv f(v) \alpha(\delta,v)}$, these shifts typically result in broadening and attenuation of the absorption line, such that the peak absorption is limited by $\alpha_0 \gamp / (k\vT\sqrt{2/\pi})$\cite{Lahad2019}. One finds however that these two shifts have opposite signs when $q>0$ in a counter-propagating configuration. The net two-photon shift has a minimum, around which different velocities contribute to the absorption at the same frequency. Notably, for a large enough detuning $\Delta \gg kv$, the light-shift is linear in velocity \mbox{${\Omega^2}/{(\Delta-kv)} \approx \Omega^2/\Delta+(\Omega^2/\Delta^2)\kp v$}, and a complete cancellation of Doppler broadening is achieved when
\begin{equation}
    {\Omega^2}/{\Delta^2}=q.
    \label{cancellation condition}
\end{equation}
 \begin{figure}
 \centering
\includegraphics[width=\columnwidth]{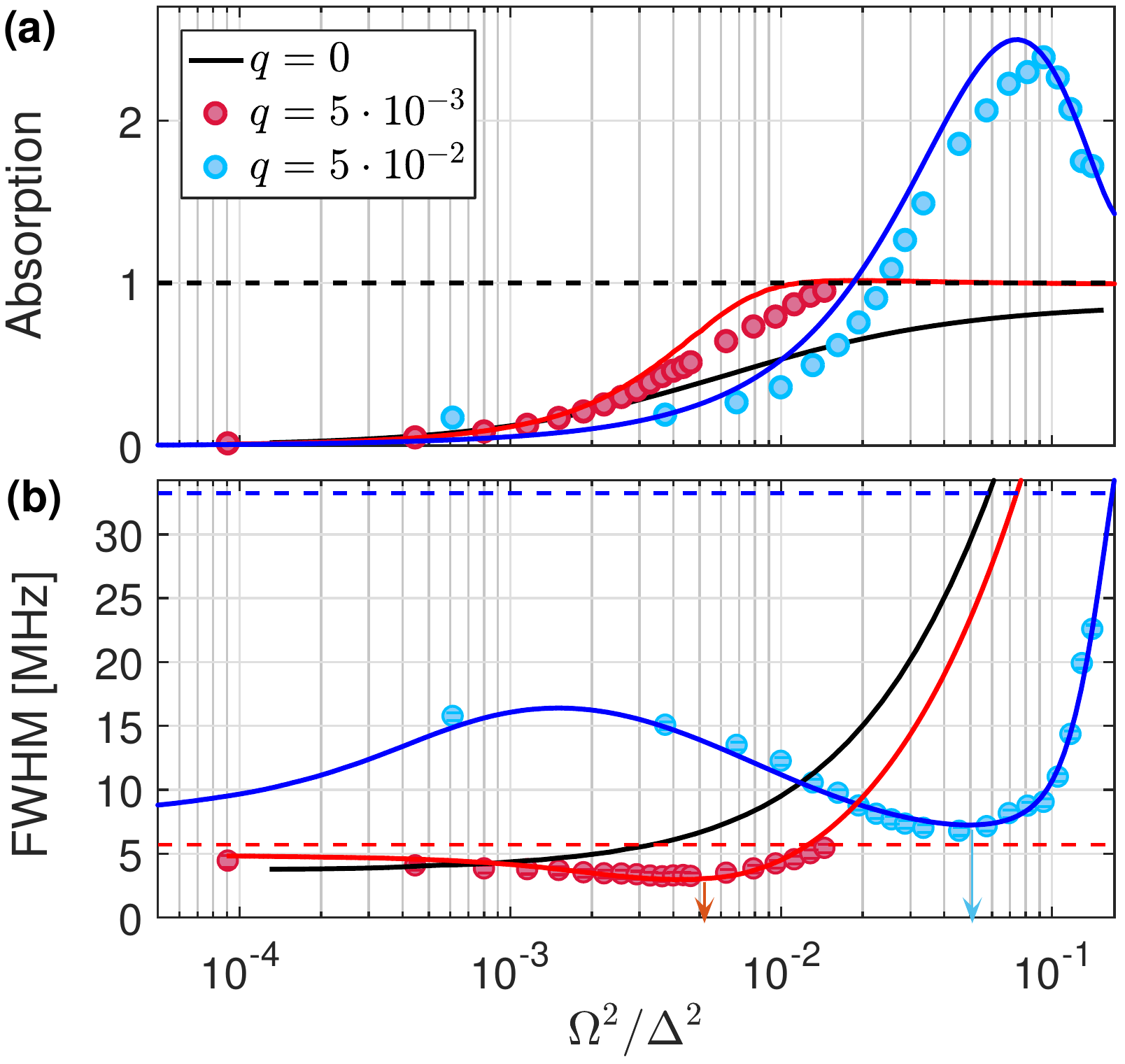}
\caption{\textbf{Power narrowing and enhancement of two-photon absorption}. (a) Measured peak probe absorption (normalized by one-photon resonant absorption in the absence of the control field) at different control powers with $q= 5\cdot 10^{-3}$ (red, measured on the $^{87}$Rb $\s \rightarrow \dstate$ transition), $q=5\cdot10^{-2}$ (blue, measured on the $\s \rightarrow \sexcited$ transition), and $q=0$ (black, calculated). Solid lines are  the result of a numerical calculation.
(b) Full width at half maximum (FWHM) for the same configurations. Both transitions are significantly narrower than the expected residual Doppler width, shown by dashed lines for $q=5\cdot 10^{-3}$ (red) and $q=5\cdot10^{-2} $ (blue). Optimal narrowing is achieved when $\Omega^2/\Delta^2=q$, marked by arrows for both configurations.  }  
\label{TPA power}
\end{figure}

\begin{figure}
\centering
\includegraphics[width=\columnwidth]{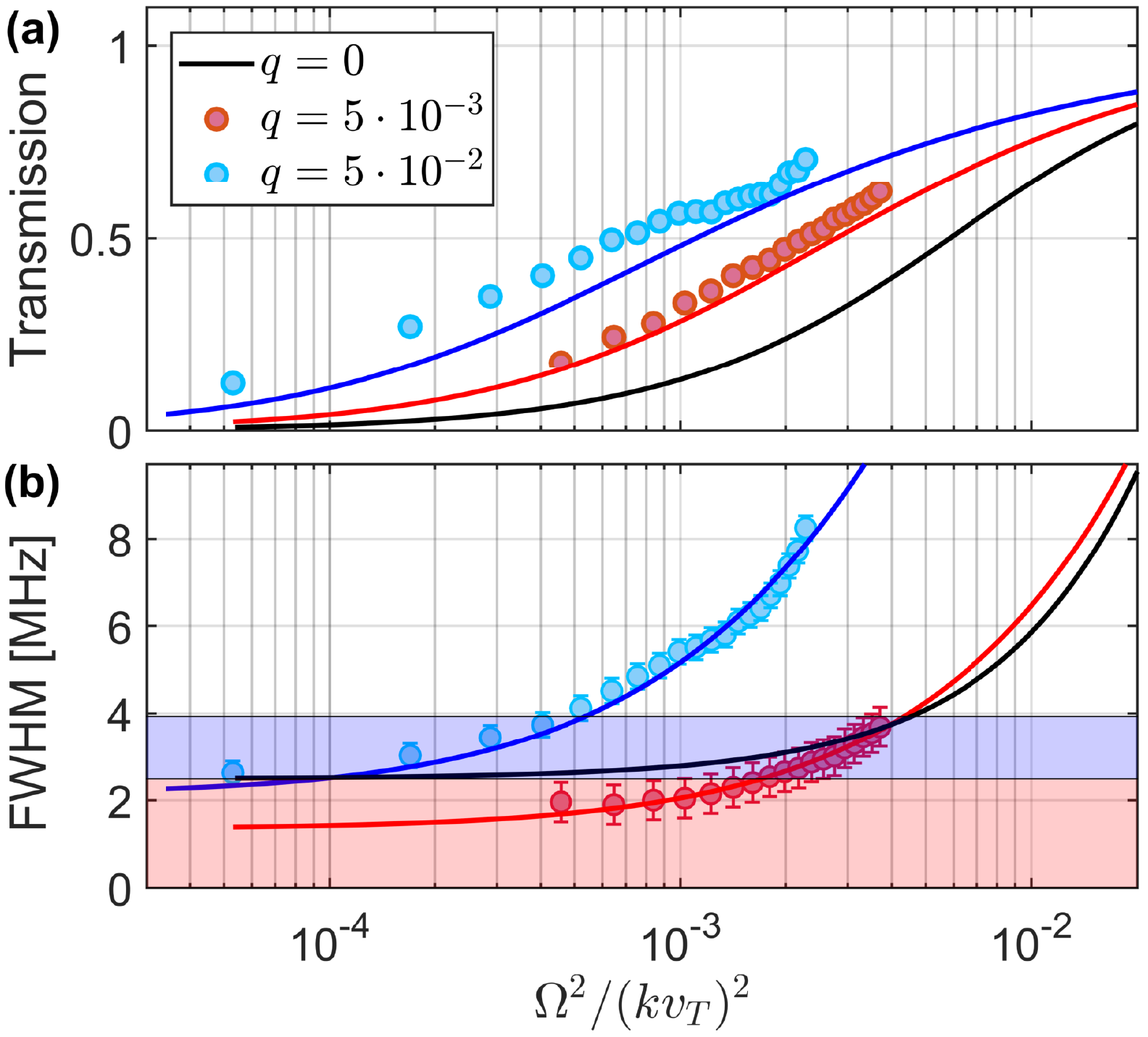} 
\caption{\textbf{Enhancement and sub-homogeneous width of EIT resonances}. (a) Measured EIT contrast at different control powers for $q= 5\cdot 10^{-3}$ (red, measured on the $\s \rightarrow \dstate$ transition), $q=5\cdot10^{-2}$ (blue, measured on the $\s \rightarrow \sexcited$ transition), and $q=0$ (black, calculated). Solid lines are the result of a numerical calculation.
(b) Full width at half maximum (FWHM) for the same configurations. The EIT lines are much narrower than the residual Doppler broadening, and, for low control powers, they are also narrower than the homogeneous width (red shading for $q=5\cdot10^{-3}$ and blue shading for $q=5\cdot10^{-2}$).}  
\label{EIT power}
\end{figure}

\emph{Experiment.---}
In our experiment, we use $^{87}$Rb with the ground level $\s$, intermediate level $\pstate$, and higher-excited levels $\dstate$ or $\sexcited$. The experiment is performed in a magnetically-shielded 10-mm long $^{87}$Rb cell at a temperature of 70 \degree C. The atoms are optically pumped to the maximally polarized ground state, such that the level structure becomes close to that of a pure three-level system.  
We measure the linewidth of the two-photon resonances as a function of control power for both two-photon absorption (Fig.~\ref{TPA power}) and EIT (Fig.~\ref{EIT power}). These measurements are done for the $\s \rightarrow \dstate $ two-photon transition, for which $\kp=2\pi/(780 $ nm) and $\kc=2\pi/(776$ nm), and for the $ \s \rightarrow \sexcited$, for which  $\kp=2\pi/(780 $ nm) and $\kc=2\pi/(741$ nm). The different transitions are thus characterized by a wavelength mismatch  $q=5\cdot10^{-3}$  and $q=5\cdot10^{-2} $, respectively. For the two-photon absorption measurements, we set the one-photon detuning to $\Delta=1180$ MHz for $q=5\cdot10^{-3}$ and to $\Delta=650$~MHz for $q=5\cdot10^{-2}$. 

In the absence of power broadening, the main decoherence mechanisms in these atomic systems are residual Doppler broadening ($q\kp\vT$; inhomogeneous), as well as radiative lifetime, laser noise, and transit-time broadening which add up to the homogeneous decoherence rate $\gamr$. In our experiment, $q\kp\vT=1.28$~MHz and $2\gamr=2.48$~MHz for $q=5\cdot10^{-3}$; $q\kp\vT=12.45$~MHz and $2\gamr=3.93$~MHz for $q=5\cdot10^{-2}$. The one-photon homogeneous decoherence rate is $2\gamp=6 $ MHz. Typically the two broadening terms combine to form so-called Voigt line-shapes, whose calculated widths are marked by dashed lines in Fig.~\ref{TPA power}(b). Nevertheless, as Figs.~(\ref{TPA power}) and (\ref{EIT power}) show, we find over a broad range of control powers that the measured linewidths of the two-photon resonances are narrower than the expected Voigt widths. These measurements agree well with full numerical calculations of the three-level susceptibility, integrated over all atomic velocities [shown as solid lines in Figs.~(\ref{TPA power}) and (\ref{EIT power})]. 
\begin{figure*}[!htb]
\centering
\includegraphics[width=1.0\textwidth]{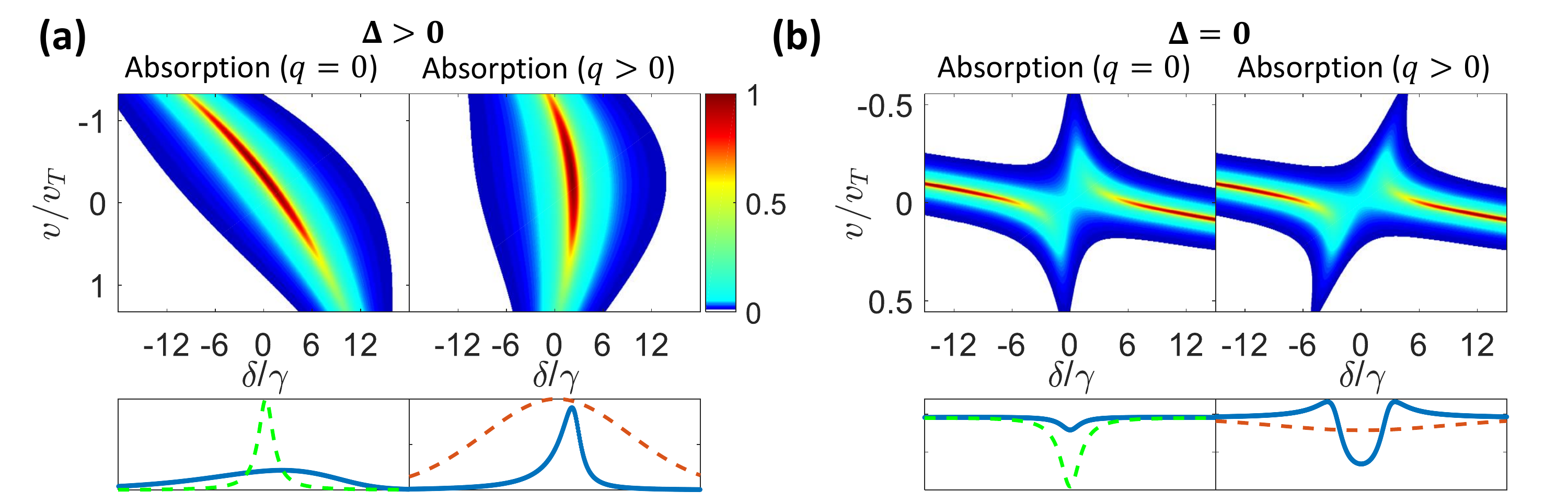} 
\caption{\textbf{Contribution of different velocity groups to the absorption spectrum $\pmb{\alpha(\delta,v)}$.} (a) Calculation of the two-photon absorption per atomic velocity group (normalized by maximal absorption) when $\Delta>0$ (here $\Delta=1.1 ~\mathrm{GHz}$). The left panels show the degenerate case $q=0$ [$\kc=\kp=2\pi/(780$ nm)], and the right panels show the two-color case $q>0$ [$\kc=2\pi/(741$ nm)]. For $q>0$, a significant fraction of the atomic velocities contribute to the absorption at the same frequency. The velocity-integrated absorption spectra (bottom, solid blue) are compared against those \textit{naively} expected due to homogeneous decoherence alone (dashed green for $q=0$) and due to residual Doppler broadening (dashed red for $q>0$). (b) The same calculations are repeated for the EIT resonance when $\Delta=0$. Here for $q>0$, the edges of the transparency window exhibit pronounced absorption peaks, which correspond to `turning points' of the maximal absorption in the velocity-detuning space.}   
\label{Fig velocity group}
\end{figure*}

For the two-photon absorption resonance (Fig.~\ref{TPA power}), we find that increasing the control power does not lead to power broadening but rather narrows the measured absorption line. The linewidth indeed reaches its minimal value around ${\Omega^2}/{\Delta^2}=q$ for both transitions, in excellent agreement with Eq. \eqref{cancellation condition}. As the Rabi frequency is increased from 0 up to this value, the resonance linewidth decreases, manifesting power narrowing. Consequently, the peak absorption increases, surpasses that calculated for $q=0$ [black curve in Fig.~2(a)], and, in the \mbox{$q=5\cdot10^{-2}$} case, even surpasses the one-photon absorption $\alpha_0 \gamp / (k\vT\sqrt{2/\pi})$. The fact that this limit is surpassed shows that power narrowing is \emph{not} a velocity-selective process, but rather engages many velocity groups in the absorption at the same detuning, thus serving as a resource for strong light-matter interaction. For a larger wavelength mismatch $q$, the mutual cancellation of the velocity-dependent light-shift and the residual Doppler shift occurs at larger Rabi frequencies, closer to the saturation of the two-photon transition, resulting in an even more striking spectral narrowing and enhancement of absorption. Thus, counter-intuitively, the wavelength mismatch $q$ is in fact a resource in this class of systems, since a larger Doppler shift can counteract the inhomogeneous light-shift at larger control powers.

For the EIT resonance (Fig.~\ref{EIT power}), we find at low control powers that the transparency window for the two-color case ($q>0$) is narrower and deeper than that expected for the degenerate case ($q=0$). In fact, in the colored areas marked in Fig.~\ref{EIT power}(b), the linewidth is even narrower than the homogeneous linewidth $2\gamr$. This sub-homogeneous feature is associated with the emergence of enhanced absorption features around the transparency line, as described by the model presented below and reproduced by a full numerical calculation.


\emph{Numerical calculations.---} 
In Fig.~\ref{Fig velocity group}(a), we compare by numerical calculations the two-photon absorption spectra of a Doppler-free system ($q=0$) with that of a two-color system ($q>0$). Naively, one may expect the former to exhibit the narrow homogeneous line (dashed green in the bottom panel), and the latter to have a broader line with width on the order of $kq\vT$ (red dashed line). In practice however, in the $q>0$ case, many velocity groups contribute to the absorption at the same frequency around the condition specified by Eq.~\eqref{cancellation condition}, whereas in the $q=0$ case, inhomogeneous light-shift broadens and attenuates the absorption line.
 
In the EIT regime $\Delta=0$, the two-photon resonances appear as transparency lines rather than absorption lines, as shown in Fig.~\ref{Fig velocity group}(b). The spectrum is comprised of many Autler-Townes doublets with different splittings and detunings, and a transparency window is formed at the narrow region where none of the velocity groups absorb light. For most velocity groups, the Doppler shift satisfies $kv\gg\gamp$, and thus Eq. \eqref{Absoprtion_theory} holds with $\Delta=0$,
\begin{equation}
\label{EIT theory}
    \alpha(\delta,v)=\alpha_0\gamp\frac{\frac{\Omega^2}{(kv)^2}\gampower}{\gampower^2+(\delta+qkv+\frac{\Omega^2}{kv})^2}.
\end{equation}
The counteracting effects of light-shift and Doppler shift result in enhanced absorption features around the central transparency window [Fig.~\ref{Fig velocity group}(b) bottom]. Furthermore, many velocities which would otherwise contribute to resonant absorption now contribute to transparency, thus enhancing the EIT contrast.



\emph{Generalization and extensions.}---
In general terms, the power narrowing mechanism we study can be understood as the counteraction of the inhomogeneous shifts of two coupled transitions. The far-detuned control field admixes each transition into the other, such that the transition to be narrowed inherits from the other transition a small fraction $\Omega^2/\Delta^2$ of its inhomogeneous shift. Power narrowing is achieved when the bare shifts of the two transitions are of opposite sign, and it is optimal when their ratio equals $-\Omega^2/\Delta^2$. This condition sets the optimal control power for a fixed $\Delta$, and it reduces to Eq.~\eqref{cancellation condition} for Doppler-broadened two-color transitions. We shall now study the sensitivity of this mechanism to the key parameters of the system and project its performance to a broader class of systems.

\begin{figure}[!tb]
\centering
\includegraphics[width=\columnwidth]{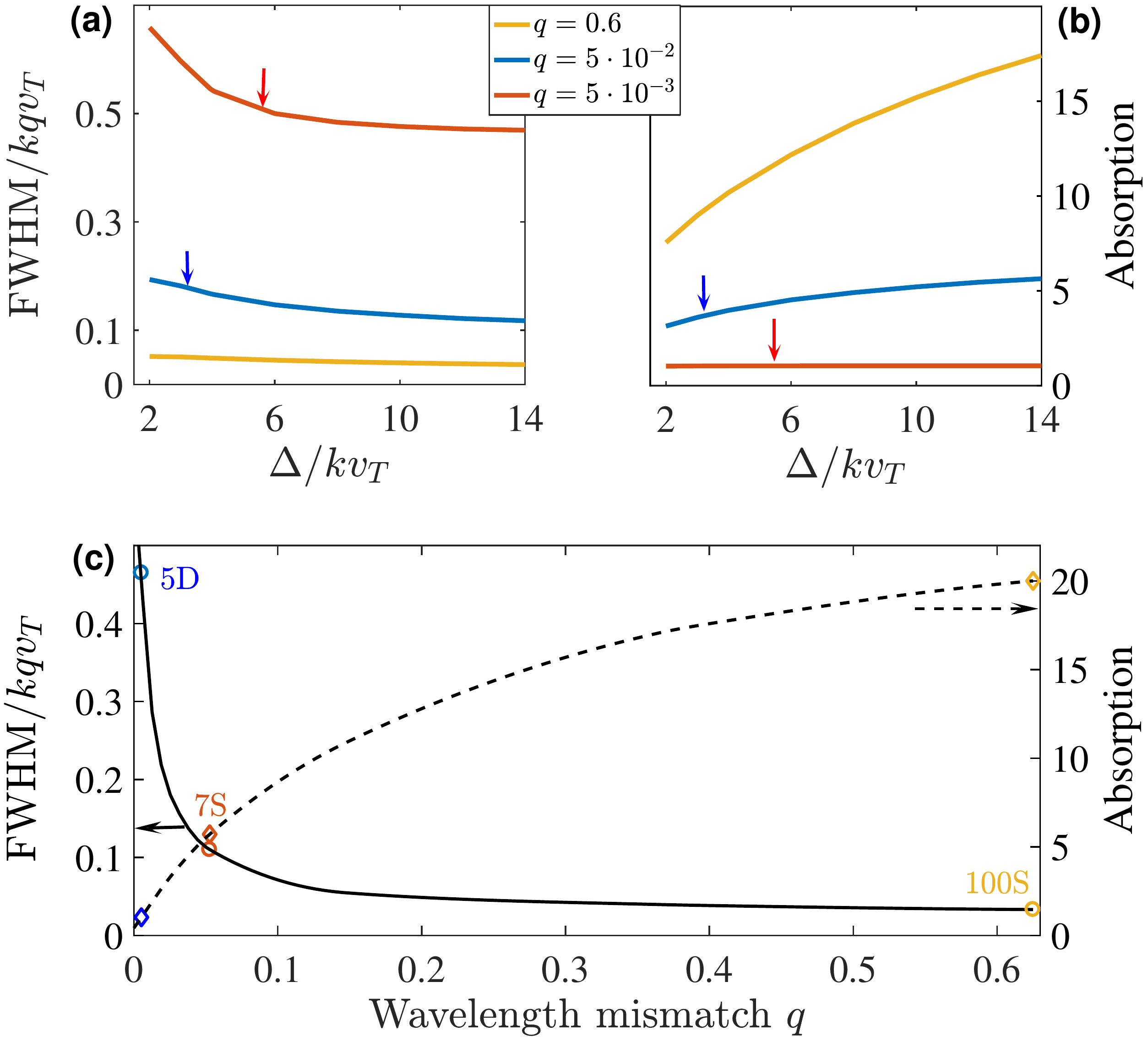}
\caption{\textbf{Dependence of power narrowing on system parameters at optimal power \boldsymbol{$\Omega^2=q\Delta^2$}:} (a) Relative narrowing (linewidth normalized by residual Doppler width) versus detuning, calculated for wavelength mismatch of $q=5\cdot10^{-3}, q=5\cdot10^{-2}$, and $q=0.6$; The latter matches transitions to high-lying Rydberg states. The working points corresponding to the experiments of Fig.~\ref{TPA power} are marked by arrows. 
(b) Gain in absorption (peak absorption normalized by one-photon absorption) versus detuning. 
(c) Dependence of power narrowing on the wavelength mismatch for fixed large detuning $\Delta=20\kp\vT$. The magnitude of the effect scales favorably with wavelength mismatch, as the counteraction of the two inhomogeneous shifts occurs at larger control powers. The points mark representative transitions in Rb.}
\label{fig5}
\end{figure}

We start by numerically verifying the optimum condition [Eq.~\eqref{cancellation condition}] for a wide range of detuning $\Delta$ and wavelength mismatch $q$. We find that indeed the narrowest line is achieved around $\Omega^2=q\Delta^2$, as long as $\Delta>\kp\vT\gg \gamp$. We then fix $\Omega^2=q\Delta^2$ and scan $\Delta$ and $q$. Figures \ref{fig5}(a) and \ref{fig5}(b) show that the line narrowing and the enhancement of absorption increase with detuning. In our experiment, the detunings were 3-5 times larger than $\kp\vT$; these points are indicated by arrows in Figs.~\ref{fig5} (a) and \ref{fig5}(b). Larger detuning would provide for more narrowing, as long as the laser power is increased accordingly. This is explained by a larger capture range of velocities, for which the weak dressing approximation is valid.

Figure~\ref{fig5}(c) shows that power narrowing scales favorably with the wavelength mismatch in two-color systems. It can therefore play an important role in systems that suffer from residual Doppler broadening that is comparable to the optical Doppler broadening. An important example for such configuration is the two-photon transition to high-lying Rydberg states in Rubidium, with $\kp = 2\pi/(780$ nm) and $\kc=2\pi/(480$ nm) \cite{firstenberg2016review}. In Fig.~\ref{fig5}, we compare this configuration ($q=0.6$) to those  experimentally confirmed in this work ($q\ll1$). As the wavelength mismatch increases, the counteraction of the two inhomogeneous shifts occurs at larger control powers, resulting in more significant narrowing. This is accompanied by enhanced absorption, approaching that of stationary atoms, which is attributed to the saturation of the two-photon transition at high powers, where $\gamp\Omega^2/\Delta^2 \gg \gamr$ for all atoms.

\begin{figure}[hbt]
\centering
\includegraphics[width=\columnwidth]{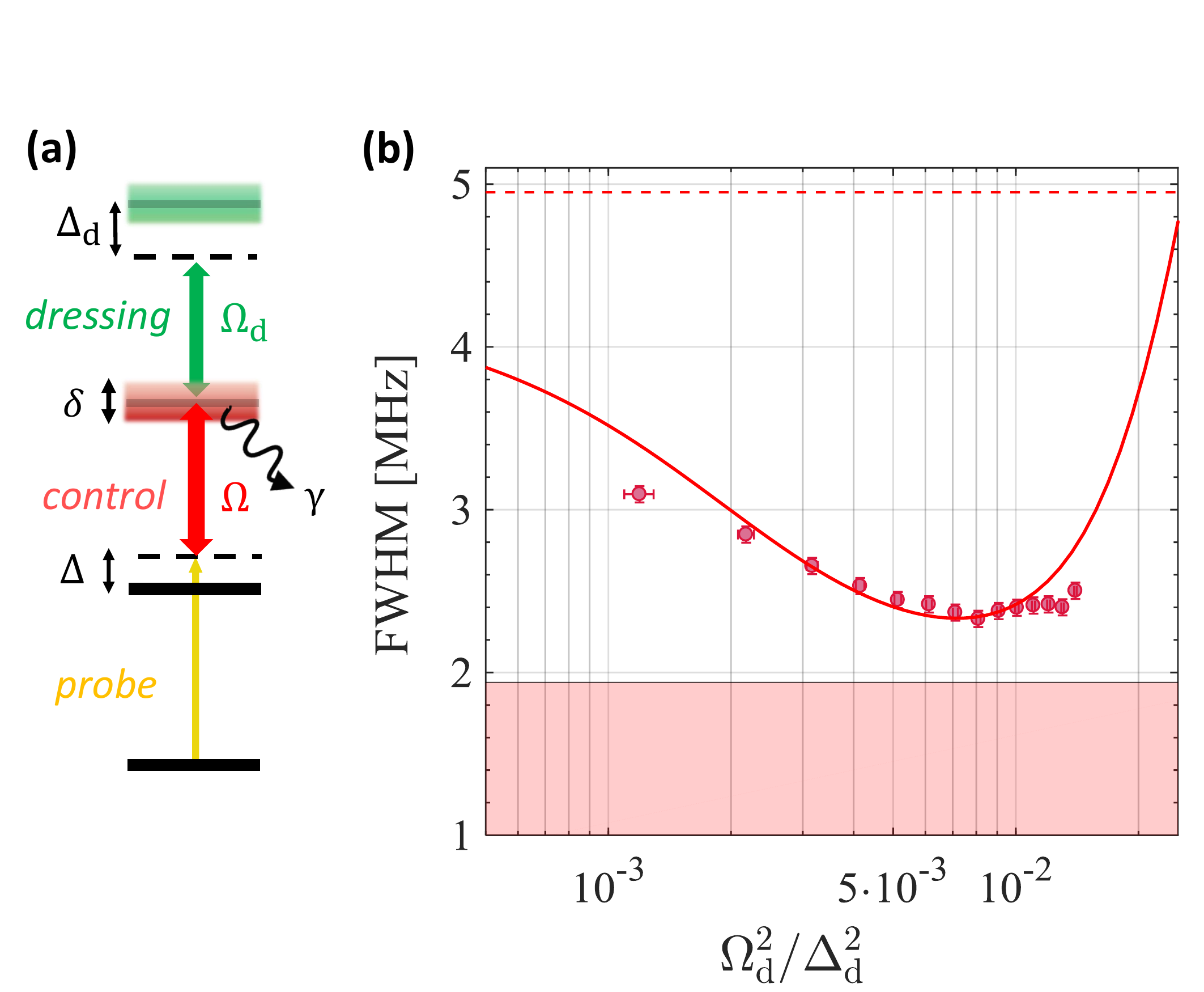}
\caption{\textbf{Power narrowing by an auxiliary dressing field.} (a) Three-color level scheme, where a dressing field with Rabi frequency $\Omega_\mathrm{d}$ is detuned by $\Delta_\mathrm{d}$ from resonance. Here $\Omega^2/\Delta^2$ is kept small, such that the two-photon transition is dominated by (residual) Doppler broadening.
(b) Measured width of the two-photon absorption line versus dressing power (circles) and corresponding numerical calculations (solid line). The width is significantly narrower than the expected residual Doppler width (dashed line) and, for optimal dressing power, reaches a value close to the homogeneous width of the two-photon transition (shaded area).
\label{1270 narrowing}
}
\label{fig6}
\end{figure}

The general picture presented above provides us with additional tools to cancel inhomogeneous broadening even when such cancellation does not occur naturally, \emph{i.e}, when Eq.~\eqref{cancellation condition} does not hold. We experimentally demonstrate how this is achieved by adding an auxiliary dressing field, as presented in Fig.~\ref{1270 narrowing}. We use the three-level system $\s\rightarrow\pstate \rightarrow\dstate$ with wavelength mismatch $q=5\cdot10^{-3}$ under weak driving, such that its two-photon linewidth is dominated by residual Doppler broadening. We add a dressing field with $k_\mathrm{d}=2\pi/(1270$ nm) detuned by $\Delta_\mathrm{d}$ = 400 MHz from the transition $\dstate \rightarrow 31F_{7/2}$, counter-propagating the control field and with Rabi frequency $\Omega_\mathrm{d}$. Since the Doppler width of the latter transition is much larger than the residual Doppler width of the $\s\rightarrow\dstate$ transition, the dressing field can be weak, $\Omega_\mathrm{d}\ll\Delta_\mathrm{d}$, ensuring negligible population transfer to the fourth level. As shown by Fig. \ref{1270 narrowing}(b), the two-photon line exhibits power narrowing with respect to the dressing power $\Omega_\mathrm{d}^2/\Delta_\mathrm{d}^2$, with the optimum obtained around $\Omega_\mathrm{d}^2/\Delta_\mathrm{d}^2=qk/k_\mathrm{d}$, similarly to Eq. \eqref{cancellation condition}.

Finally, various non-idealities may limit the magnitude of the effect presented in this work. First, for a given detuning $\Delta$ and wavelength mismatch $q$, the magnitude of power narrowing and the absorption enhancement are bounded by the ratio of inhomogeneous to homogeneous broadening. In the experimental results presented here, terms such as laser linewidth and transit-time broadening constitute a significant fraction of the homogeneous width, on top of the natural lifetime. Improving these factors would further increase the achievable power narrowing. Second, in the absence of optical pumping, the manifold nature of the atomic levels might reduce the overall effectiveness of the narrowing effect, as the optimal condition of Eq. \eqref{cancellation condition} cannot be simultaneously met for several transitions with different coupling strengths. Nevertheless, for the entire manifold, the Doppler shift and the light-shift are of opposite signs and counteract each other over a broad range of laser powers. Thus, while its magnitude is reduced, power narrowing still occurs for degenerate multi-level systems, even in the presence of non-idealities such as non-optimal optical pumping.

\emph{Conclusion.---} We have identified a class of systems where velocity-dependent light-shift and Doppler shift can counteract each other. For low driving powers, the two-photon absorption-line becomes narrower with increasing power, up to the point of complete cancellation of Doppler broadening, after which the line broadens as typically occurs with power broadening. An important distinction between the power narrowing mechanism and other spectral-narrowing phenomena is that power narrowing engages many velocity groups in the resonant process and is, in principle, not velocity selective.
While power narrowing of the two-photon absorption resonance results in an enhanced absorption (thus enabling stronger light-matter interaction), a similar mechanism in the EIT resonance results in enhanced transparency contrast and narrow linewidth. These features would lead to a smaller group-velocity of light pulses propagating through such media. We have found that optimal power narrowing in two-photon absorption is always achieved when the ratio between the inhomogeneous widths of the two-photon and one-photon transitions equals $\Omega^2/\Delta^2$. The latter can be understood in a dressed state picture as a mutual cancellation of the inhomogeneous broadenings of the constituent bare levels. We have demonstrated how this picture is applied in other configurations, by using an auxiliary dressing field to cancel the inhomogeneous broadening of a system where such cancellation does not occur naturally. In all configurations, we found the qualitative effect to be robust to the employed dressing powers and detunings. 

While we have studied here the cancellation of Doppler broadening in continuous wave spectroscopy, akin to Rabi spectroscopy, the same rephasing mechanism should also apply to pulsed schemes, akin to Ramsey spectroscopy. Importantly, one could eliminate Doppler dephasing of collective excitations in pulsed light-storage schemes by applying a rephasing power-narrowing field during storage.

These results establish power narrowing as an important tool for quantum technologies based on strong and coherent light-matter interaction at ambient conditions, where inhomogeneous broadening is often the limiting factor. 

 \emph{Acknowledgements---} We acknowledge financial support by the Israel Science Foundation and ICORE, the European Research Council starting investigator grant Q-PHOTONICS 678674, the Pazy Foundation, the Minerva Foundation with funding from the Federal German Ministry for Education and Research, and the Laboratory in Memory of Leon and Blacky Broder.



\bibliography{Ladder_Spec_BIB}

\end{document}